\documentclass[sigconf]{acmart}
\usepackage{threeparttable}
\usepackage{makecell}
\usepackage{multirow}
\usepackage{diagbox}
\usepackage{algorithm}
\usepackage{algorithmic}
\usepackage{setspace}
\usepackage[section]{placeins}
\AtBeginDocument{%
  \providecommand\BibTeX{{%
    \normalfont B\kern-0.5em{\scshape i\kern-0.25em b}\kern-0.8em\TeX}}}

\setcopyright{acmcopyright}
\copyrightyear{2018}
\acmYear{2018}
\acmDOI{XXXXXXX.XXXXXXX}

\acmConference[Conference acronym 'XX]{Make sure to enter the correct
  conference title from your rights confirmation emai}{June 03--05,
  2018}{Woodstock, NY}
\acmPrice{15.00}
\acmISBN{978-1-4503-XXXX-X/18/06}

\begin{document}

\title{A New Creative Generation Pipeline for Click-Through Rate with Stable Diffusion Model}
\author{Hao Yang}
%\orcid{}
\affiliation{%
  \institution{Shopee Discovery Ads}
  \city{Beijing}
  \country{China}
  }
%\email{xxx@shopee.com}
\author{Jianxin Yuan}
%\orcid{}
\affiliation{%
  \institution{Shopee Discovery Ads}
  \city{Beijing}
  \country{China}
  }
%\email{xxx@shopee.com}
\author{Shuai Yang}
%\orcid{}
\affiliation{%
  \institution{Shopee Discovery Ads}
  \city{Beijing}
  \country{China}
  }
%\email{xxx@shopee.com}
\author{Linhe Xu}
%\orcid{}
\affiliation{%
  \institution{Shopee Discovery Ads}
  \city{Beijing}
  \country{China}
  }
%\email{xxx@shopee.com}
\author{Shuo Yuan}
%\orcid{}
\affiliation{%
  \institution{Shopee Discovery Ads}
  \city{Beijing}
  \country{China}
  }
%\email{xxx@shopee.com}
\author{Yifan Zeng}
%\orcid{}
\affiliation{%
  \institution{Shopee Discovery Ads}
  \city{Beijing}
  \country{China}
  }
%\email{xxx@shopee.com}

\begin{abstract}
In online advertising scenario, sellers often create multiple creatives to provide comprehensive demonstrations, making it essential to present the most appealing design to maximize the Click-Through Rate (CTR). However, sellers generally struggle to consider users' preferences for creative design, leading to the relatively lower aesthetics and quantities compared to Artificial Intelligence (AI)-based approaches. Traditional AI-based approaches still face the same problem of not considering user information while having limited aesthetic knowledge from designers. In fact that fusing the user information, the generated creatives can be more attractive because different users may have different preferences. To optimize the results, the generated creatives in traditional methods are then ranked by another module named creative ranking model. The ranking model can predict the CTR score for each creative considering user features. However, the two above stages (generating creatives and ranking creatives) are regarded as two different tasks and are optimized separately. Specifically, generating creatives in the first stage without considering the target of improving CTR task may generate several creatives with poor quality, leading to dilute online impressions and directly making bad effectiveness on online results.

In this paper, we proposed a new automated \textbf{C}reative \textbf{G}eneration pipeline for \textbf{C}lick-\textbf{T}hrough \textbf{R}ate (\textbf{CG4CTR}) \footnote{The code is at \href{https://github.com/HaoYang0123/Creative\_Generation\_Pipeline}{https://github.com/HaoYang0123/Creative\_Generation\_Pipeline}.} with the goal of improving CTR during the creative generation stage. In this pipeline, a new creative is automatically generated and selected by stable diffusion method with the LoRA model and two novel models: prompt model and reward model. Our contributions have four parts: 1) The inpainting mode in stable diffusion method is firstly applied to creative image generation task in online advertising scene. A self-cyclic generation pipeline is proposed to ensure the convergence of training. 2) Prompt model is designed to generate individualized creative images for different user groups, which can further improve the diversity and quality of the generated creatives. 3) Reward model comprehensively considers the multi-modal features of image and text to improve the effectiveness of creative ranking task, and it is also critical in self-cyclic generation pipeline. 4) The significant benefits obtained in online and offline experiments verify the significance of our proposed method.

% we achieve a breakthrough by introducing LoRA, along with two novel modules: the Prompt Model (PM) and the Reward Model (RM). 
% Finally, in online A/B test experiments, we obtain 10.4\% of CTR improvements after uploading the AI-generated creatives compared with the original creatives designed by sellers.
%% , self-circulating process is raised with iterative updating LM and PM to 
% We achieve a breakthrough by introducing LoRA, along with two novel modules: the Prompt Model and the Reward Model. This innovative pipeline enables the generation of personalized creatives that demonstrate enhanced CTR performance. We leverage the self-cycling training capability to significantly accelerate the convergence speed of the model, thereby substantially reducing the exploration cost associated with generating individual creatives.
%LoRA is responsible for creative generation given one suitable prompt which is selected by PM. RM is responsible for judging the good or bad of creatives to facilitate the subsequent iteration.

\end{abstract}




\maketitle

\vspace{-10pt}
\section{INTRODUCTION}
% In the e-commerce recommendation scenario, improving Click-Through Rate (CTR) is very important. Creative, as the image showcasing the details of product, plays a crucial role in attracting user clicks and improving CTR. Then lots of related works are appeared. Next, we will introduce the previous work into two parts: text-to-image (T2I) generation task in AI-Generated Content (AIGC) and creative generation methods in e-commerce scenario.
Recently, the topic of AI-Generated Content (AIGC) has developed rapidly, such as GLIDE \cite{nichol2021glide}, DALL-E 1-3 \cite{ramesh2021zero, ramesh2022hierarchical, betker2023improving}, Imagen \cite{saharia2022photorealistic}, Stable Diffusion (SD) \cite{rombach2022high} and ChatGPT \cite{ouyang2022training}. One of the most representative tasks is text-to-image (T2I) generation, which has received much attention in both academia and industry. The target of T2I task is to generate images with similar semantics given input language prompt. Currently, the diffusion-based methods \cite{dhariwal2021diffusion, ho2020denoising, rombach2022high} have become the state-of-the-art (SoTA) method in T2I task due to the stationary training objective and easy scalability. Although these diffusion methods have a satisfactory performance on the T2I task, directly using these methods in creative image generation is infeasible in the advertising scene because they will modify the main product image. Traditional creative generation methods in the advertising scene use deep learning to generate some objects/tags \cite{vempati2020enabling, mishra2020learning}, dense captions \cite{gao2022caponimage} or layout information \cite{zhou2022composition, inoue2023layoutdm, qu2023layoutllm} on image. Different from these methods, we first use diffusion model to generate background images while keeping the main product information unchanged in creative generation task for the advertising scene. In the experimental analysis, we found that modifying the background while keeping the visual pixels of the main product unchanged can also significantly improve the CTR.

\begin{figure}[h]
\setlength{\abovecaptionskip}{0.0cm}
\setlength{\belowcaptionskip}{-0.2cm}
  \centering
  \includegraphics[width=1.0\linewidth]{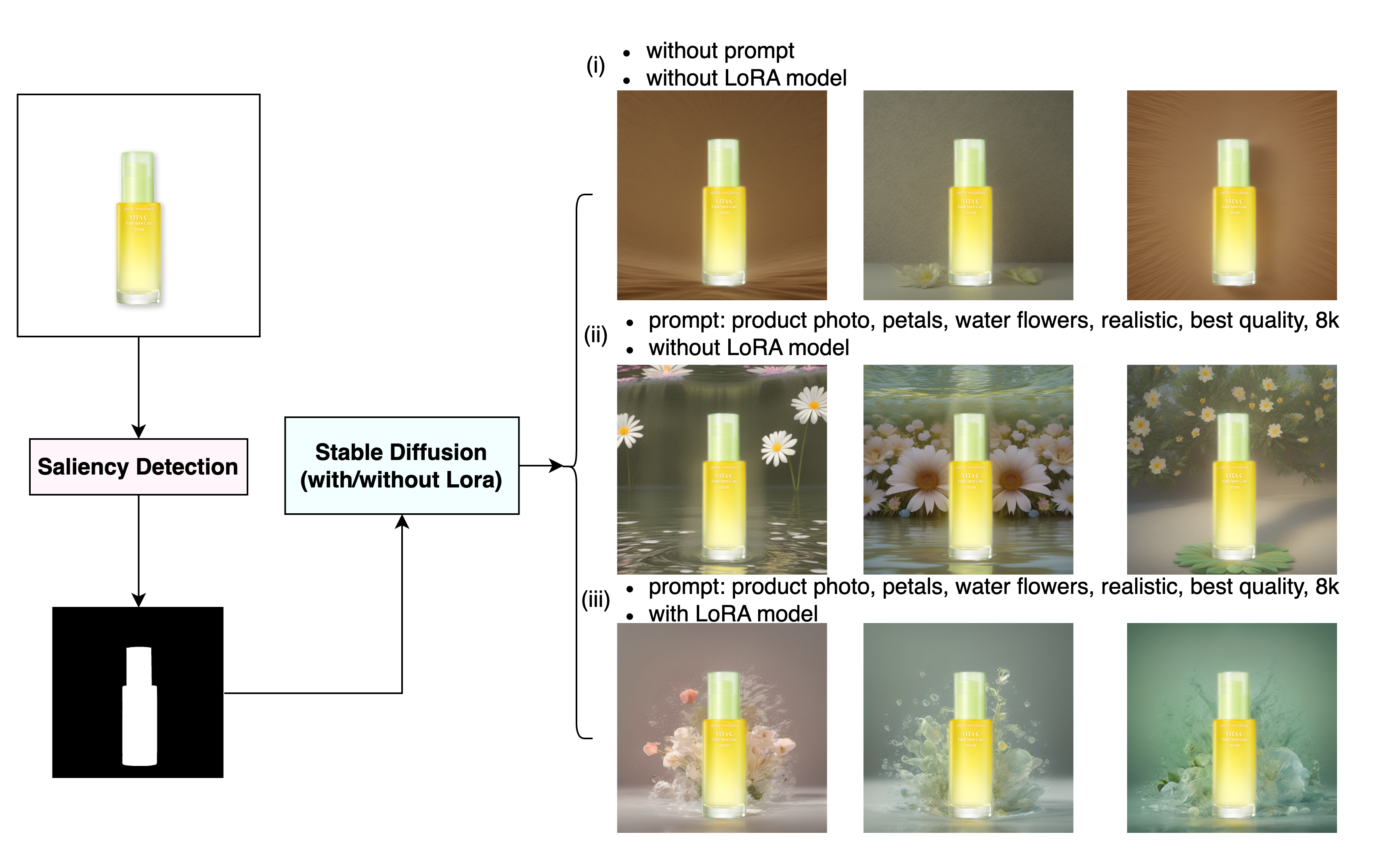}
  \caption{Generating creatives only by modifying background using SD method in inpainting mode.}
  \label{fig-sd-workflow}
  \Description{Generating creatives only by modifying background using SD method in inpainting mode.}
  % \vspace{-10pt}
\end{figure}

In addition, when using the diffusion model, there are two very important factors, prompt and diffusion models, that will directly affect the effect of the generated image. As shown in Fig. \ref{fig-sd-workflow}, for the original product image uploaded by the seller, the background image is simple with solid white color. To make the background more attractive, we can use the inpainting mode in SD method to generate more individualized background images with richer colors. However, if a proper prompt (a prompt consists of several tokens) is not added into the generation process, the result is often poor with the cluttered background. Then we need to use a Prompt Model (PM) to select a good prompt, which is then added to guide the generation process. Despite that a good background can be generated given an acceptable prompt, the diffusion model is not fine-tuned in our industrial data, and the effect is not optimal. It is necessary to fine-tune diffusion model in our data. However, the parameters of diffusion model are very huge, and directly training whole model is not realistic because of high training cost. Therefore, Low-Rank Adaptation (LoRA) mechanism \cite{ryu2023low, hu2021lora} is used to speed up the training process.

\vspace{-15pt}

As described earlier, Prompt Model (PM) is important to guide image generation. PM considers user feature (\emph{e.g.}, user age, user gender), item feature (\emph{e.g.}, item ID and category ID) and candidate tokens as input, then outputs score for each input token. We select top-$p$ tokens to assemble into one prompt. Considering that different users have different preferences for creative images, for example, young users are more inclined to cool color and electronic style while old users are more inclined to warm color and minimalistic style. Then user feature is considered as a vital feature to generate individualized prompts for different user groups. The traditional creative generation methods \cite{vempati2020enabling, mishra2020learning, gao2022caponimage, zhou2022composition, inoue2023layoutdm, qu2023layoutllm} do not consider the user information in generating creative image process. Some works \cite{zhao2019you, lin2022joint, wang2021hybrid, you2023image} use another model to rank all generated creatives considering user information and select the best one to show online. These traditional pipelines do not maximize the online performance because in first stage (generating creative stage), they do not regard the improving CTR as the target, resulting in many "bad" creatives diluting the online impressions. So it is necessary and critical to directly generate more attractive creatives by considering user information and regarding the CTR as an optimization target in generating stage. Currently, several works \cite{tang2023zeroth, black2023training} focus on enhancing the generation quality of diffusion models by reinforcement learning with human feedback. Although these methods can generate creatives using diffusion models for a specific target, such as CTR, they still face the problems of modifying the original product with relatively low efficiency and diversity.

To guide the prompt model and LoRA model to generate more attractive creatives, another Reward Model (RM) is used to predict the Click-Through Rate (CTR) score for each creative image given one product. For each product, if too many creatives are uploaded online, it will lead to sub-optimal performance due to the inferior performance of the generated creatives \cite{wang2021hybrid}. Then we need RM to select the proper creatives for displaying online. At the same time, RM can be used to generate samples (good creatives with high predicted scores and bad creatives with low scores) to train LoRA and prompt models.

To summarize, there are four contributions in this work: 1) The inpainting mode in stable diffusion method is firstly applied to creative image generation task in online advertising scene. A self-cyclic generation pipeline is proposed to ensure the convergence of training. 2) Prompt model is designed to generate individualized creative images for different user groups, which can further improve the diversity and quality of the generated creatives. 3) Reward model comprehensively considers the multi-modal features of image and text to improve the effectiveness of creative ranking task, and it is also critical in self-cyclic generation pipeline. 4) The significant benefits obtained in online and offline experiments verify the significance of our proposed method.  % RM: achieve state-of-the-art (SoTA) effect in creative ranking task

\vspace{-10pt}
\section{RELATED WORKS}
%In this section, we present a summary of related works in two parts: text-to-image generation and creative CTR prediction.
\subsection{Text-to-image Generation}
Recently, in text-to-image generation (T2I) task, more and more deep learning models to generate image given text are emerging. For example, the generative adversarial networks (GANs) \cite{goodfellow2014generative, reed2016generative} use a generator and a discriminator to generate image, where generator network is responsible for generating image and discriminator network is to distinguish the generated image and the real one. There are some related GANs works, such as ProGAN \cite{karras2017progressive}, StyleGAN \cite{karras2019style, karras2020analyzing, karras2021alias}, Projected GAN \cite{sauer2021projected}, VQGAN \cite{esser2021taming}. Different from GANs, some autoencoder (AE) related works appear, such as deep autoencoder \cite{hinton2006reducing}, variational autoencoder (VAE) \cite{kingma2013auto}, vector quantised-variational autoencoders (VQ-VAE) \cite{van2017neural, razavi2019generating}. The encoder receives the input and encodes it in a latent space of a lower dimension while the decoder decodes this vector to produce the original input.

As auto-regressive models developed in text generation recently, a lot of works adopted this to achieve amazing results for T2I generation, such as DALL-E 1-3 \cite{ramesh2021zero, ramesh2022hierarchical, betker2023improving}, CogView \cite{ding2021cogview} and Pariti \cite{yu2022scaling}. Currently, the diffusion models (DM) \cite{nichol2021glide, rombach2022high, saharia2022photorealistic} appeared as SoTA T2I methods due to the natural fit to inductive biases from image data. For example, GLIDE \cite{nichol2021glide} uses an adequate text-guidance strategy to generate and edit photorealistic image. Latent DM \cite{rombach2022high} enables DM training on limited computational resources while retaining the quality and flexibility by applying in latent space of powerful pre-trained autoencoders. Imagen \cite{saharia2022photorealistic} directly diffuses pixels using a pyramid structure without using latent images.

Recently, some works \cite{vempati2020enabling, mishra2020learning, gao2022caponimage, wang2022creagan, zhou2022composition, inoue2023layoutdm, qu2023layoutllm} study the task of creative generation in the advertising scene. For example, \cite{vempati2020enabling} automatically annotates the objects from the image and generates optimal banner layout information. CapOnImage \cite{gao2022caponimage} generates dense captions at different locations of the image by pre-training and fine-tuning a multi-modal framework. CreaGAN \cite{wang2022creagan} generalizes to other product materials by utilizing existing design elements (\emph{e.g.}, background material). Other works \cite{zhou2022composition, inoue2023layoutdm, qu2023layoutllm} generate layout information of some elements, such as text and logo, given the original image. % For example, LayoutLLM-T2I \cite{qu2023layoutllm} fuses the layout planning and image generation into one pipeline by using diffusion method, which can improve the quality of generated images when importing layout information.

Different from the above methods, we propose a novel pipeline for generating creatives in advertising scene. A stable diffusion method is used in inpainting mode to generate only the background pixels, while keeping the other pixels on the main product unchanged. On the other hand, by introducing user information, we can support individualized generation for various user group.

\vspace{-17pt}
\subsection{Creative CTR Prediction}
Creative CTR prediction task is to rank all creative images provided by sellers or automatically generated by models, given one product or advertisement. AMS \cite{ge2018image} combines the behavior images into modeling user preference in CTR prediction. PEAC \cite{zhao2019you} evaluates and selects advertisement creatives offline before being placed online. VAM \cite{wang2021hybrid} uses a visual-aware ranking model to order the creatives according to their visual appearance. CACS \cite{lin2022joint} considers the different image features in ranking stage to joint optimization of intra-advertisement creative selection and inter-advertisement ranking. CLIK \cite{ko2022contrastive} considers the topic of the given product to select representatives from several images with intuitive visual information by contrastive learning. \cite{you2023image} proposes a two-stage pipeline method with content-based recall model and CTR-based ranking model to find an appropriate image. Compared to these methods, our proposed reward model performs better because of the multi-modal structure and pre-training techniques.

\vspace{-5pt}
\section{METHODS}

\subsection{Pipeline}
The pipeline is shown in Fig. \ref{fig-pipeline} and the pseudo-code is shown in Alg. \ref{alg1}. In the creative generation pipeline, for each item, we first select one original image with a relatively clean background provided by sellers, which then is inputted into saliency model \cite{qin2022highly} to get the saliency object image and mask image. We use Stable Diffusion (SD) \cite{rombach2022high} method to generate the creatives given the saliency image, LoRA model \cite{ryu2023low, hu2021lora} and prompts where LoRA model is fine-tuned in our commercial data and prompts are selected by another prompt model trained on online creative clicked data. The LoRA model and prompt model are critical for the impact on generating creatives, especially since different prompts can cause creatives to vary greatly. To consistently generate good creatives, we need to fine-tune LoRA model and prompt model in this pipeline. The details are described in Section 3.2 and 3.3. Next, we use reward model (details in Section 3.4) to predict the CTR score for each generated creative, and only top-\emph{k} creatives are retained to further train LoRA model while these top-\emph{k} creatives are treated as positive samples and others are treated as negative samples to train prompt model.

Last but not least, various users may have distinct preferences on creatives. So we need to generate different types of creatives for various users. Then user information is considered in prompt model, although the input product or image is the same, the prompts may need to be adjusted based on the characteristics of the user group, and the resulting generated creatives can also be different. To improve the quality and stability of the generated creative, the LoRA model and prompt model are updated iteratively. Precisely, in first step, only parameters in LoRA model are updated while parameters in prompt model are fixed; in second step, parameters in LoRA model are fixed while parameters in prompt model are updated, ensuring the convergence of the entire framework.

\begin{algorithm}
\small
\setlength{\belowcaptionskip}{-0.5cm}
	%\textsl{}\setstretch{1.0}
        %\renewcommand{\baselinestretch}{0.2}
        \setlength{\baselineskip}{8pt}
	\renewcommand{\algorithmicrequire}{\textbf{Input:}}
	\renewcommand{\algorithmicensure}{\textbf{Output:}}
	\caption{Creative Generation Pipeline for CTR.}
	\label{alg1}
	\begin{algorithmic}[1]
            \REQUIRE
            $M$ is the number of items \\
            $N$ is the number of training epochs \\
            $\theta_{p}$ is the parameters of prompt model \\
            $\theta_{l}$ is the parameters of LoRA model \\
            $\theta_{r}$ is the parameters of reward model \\
            $\theta_{s}$ is the parameters of saliency model \\
            $\theta_{c}$ is the parameters of CLIP-Interrogator model \\
            $p$ is the number of tokens select by prompt model \\
            $q$ is the number of generated creative images \\
            $k$ is the number of images select by reward model \\
		\STATE Initialization: $\theta_{p}$ and $\theta_{l}$ are trained on clicked data as the model parameters for initialization. Note that $\theta_{r}$ has been trained beforehand while $\theta_{s}$ and $\theta_{c}$ are the public models, these three model parameters are frozen in this pipeline.
            \FOR{$i = 1 \to N$} 
                \STATE{G, B = $\emptyset$, $\emptyset$, which refer to the set of good case and bad case, respectively.}
                \FOR{$j = 1 \to M$}
                    \STATE{Get the original images of $item_j$ provided by seller}
                    \STATE{Get saliency images and masked images by $\theta_{s}$}
                    \STATE{Select top-$p$ tokens as the appropriate
 prompt $prompt_j$ for $item_j$ by $\theta_{p}$ given user group feature}
                    \STATE{Generate $q$ creative images for $item_j$ by $\theta_{l}$ given $prompt_j$}
                    \STATE{Sort images by predicted CTR scores from $\theta_{r}$}
                    \STATE{Select top-$k$ images as positive samples and other images as negative samples}
                    \STATE{Generate the prompt for each image by $\theta_{c}$}
                    \STATE{Insert samples of ($item_j$, $image_1$, $prompt_1$, $score_1$), ... ($item_j$, $image_k$, $prompt_k$, $score_k$) into G}
                    \STATE{Insert samples of ($item_j$, $image_{k+1}$, $prompt_{k+1}$, $score_{k+1}$), ... ($item_j$, $image_q$, $prompt_q$, $score_q$) into B}
                \ENDFOR
                \IF{$i \; \% \; 2 == 0$}
                    \STATE{Train prompt model $\theta_{p}$ given G and B by Equation \ref{eq-promptloss}}
                \ELSE
                    \STATE{Train LoRA model $\theta_{l}$ given G by Equation \ref{eq-loss-sd-inp}}
                \ENDIF
            \ENDFOR
		\ENSURE $\theta_{p}$ and $\theta_{l}$
	\end{algorithmic}  
\end{algorithm}

\vspace{-8pt}
\begin{figure*}
\setlength{\abovecaptionskip}{-0.2cm}
\setlength{\belowcaptionskip}{-0.4cm}
  \centering
  \includegraphics[width=0.8\textwidth]{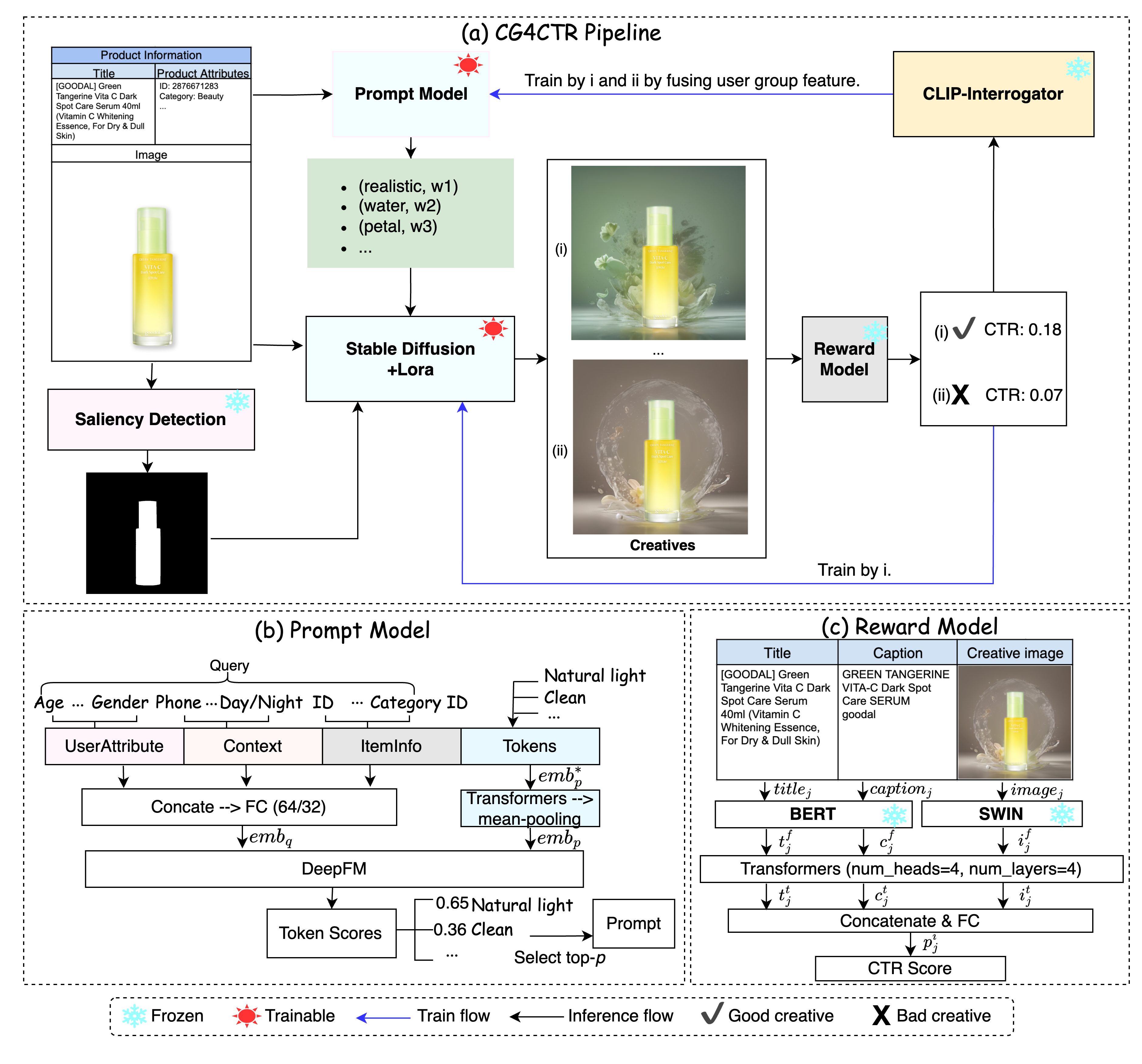}
  \caption{(a) Creative generation pipeline. (b) The structure of prompt model. (c) The structure of reward model.}
  \label{fig-pipeline}
  \Description{(a) Creative generation pipeline. (b) The structure of prompt model. (c) The structure of reward model.}
  \vspace{-2pt}
\end{figure*}

\vspace{-7pt}
\subsection{LoRA Model in Stable Diffusion}
As the original creatives designed by sellers may have quality and quantity problems, we need to modify the images to a certain extent. Recently, a lot of works \cite{ho2020denoising, nichol2021glide, ramesh2022hierarchical, li2023gligen} use diffusion models to generate images, which uses a variational autoencoder (VAE) \cite{kingma2013auto} to operate the diffusion process in a low-dimensional latent space with good computational efficiency. In detail, one image \emph{I} is given into an encoder $VAE$ to get the latent feature $Z$, \emph{i.e.}, $Z = VAE(I)$. A diffusion model is to denoise an input latent feature $Z_t$ at each timestep $t$ conditioned on a given prompt. During the training process, for each timestep $t$, the diffusion denoising network $\varepsilon_\theta$ is optimized to remove the noise $\varepsilon$ from the noised version of latent code $Z_t$ conditioned on prompt $y$ shown in following equation, where $\tau_\theta$ is the CLIP \cite{radford2021learning} text encoder, $y$ is the given text or prompt, $\varepsilon$ is the noise sampled according to a standard normal distribution. The diffusion network $\varepsilon_\theta$ often uses UNet \cite{ronneberger2015u} consisting of convolution, self-attention and cross-attention layers. $Z_t$ is the noised embedding encoded by a scheduler in low-dimensional space with VAE model given image $I$ and noise $\varepsilon$. The scheduler is used to add noise information into low-dimensional embedding from the given image.
\begin{small}
\begin{equation}
    L = \mathbb{E}_{Z \sim VAE(I), \; y, \; \varepsilon \sim N(0,1), \; t} \left[ ||\varepsilon - \varepsilon_\theta(Z_t, t, \tau_\theta(y))||_2^2 \right] \;, where \label{eq-loss-sd}
\end{equation}
\begin{equation}
    Z_t = scheduler(VAE(I), \; t, \; \varepsilon) \label{eq-loss-vae}
\end{equation}
\end{small}

However, the standard SD method cannot be used in our pipeline because it will modify the main product in advertising platform. For example, the product image is a pair of shoes uploaded by a seller, after the SD generating process, the generated image might be a pair of socks or a hat, which is not accepted. To generate one image with a higher quality while leaving the main content of original product unchanged, we can get the pixels (referred as $I_{sal}$) of the main product image $I$ by using the saliency detection network \cite{qin2022highly}. The pre-trained saliency detection network can point out the main product locations and then $I_{sal}$ are masked and only background pixels $I_{bkg}$ are generated by SD method in inpainting mode \cite{suvorov2022resolution}. The only difference between the inpainting mode and normal mode in SD is that the inpainting mode concatenates the low-dimensional embedding ($VAE(I_{sal})$) encoded by VAE given masked pixels (Equation \ref{eq-zt-inp}). The training loss is similar to the normal mode.
\vspace{-5pt}
\begin{small}
\begin{equation}
    L = \mathbb{E}_{Z' \sim VAE(I, \; I_{sal}), \; y, \; \varepsilon \sim N(0,1), \; t} \left[ ||\varepsilon - \varepsilon_\theta(Z_t', t, \tau_\theta(y))||_2^2 \right] \;, where \label{eq-loss-sd-inp}
\end{equation}
\begin{equation}
    Z_t' = concat(scheduler(VAE(I), \; t, \; \varepsilon), \; VAE(I_{sal}))
\label{eq-zt-inp}
\end{equation}
\end{small}

Because of the difference in image distribution between our commercial data and public data, we should fine-tune the public SD model on our commercial data. However, it is very difficult to fine-tune SD model because of huge number of parameters in SD model, including UNet \cite{ronneberger2015u} and CLIP \cite{radford2021learning}. Low-Rank Adaptation (LoRA) \cite{ryu2023low, hu2021lora} can help to solve the above problem, which allows us to use low-rank adaptation technology to quickly fine-tune SD model. With fine-tuned LoRA model, new creative image with a generated background can be acquired by SD method given an appropriate prompt. 

We take Fig. \ref{fig-sd-workflow} to explain the importance of the prompt and LoRA in generating creative workflow. The original image of the product given by seller is poor due to the simple colored background. To improve the quality of this creative, firstly we get masked pixels of the main product (\emph{e.g.}, skin care product) by saliency detection model \cite{qin2022highly}. Given the original image and masked image, we use SD method in inpainting mode to get better creatives with suitable backgrounds. Of course, to explain the importance of prompt and LoRA model in generating process, firstly we only use the SD method without prompt and LoRA model, the results are not as expected with mussy background same as the original image (three upper images in Fig. \ref{fig-sd-workflow}). To improve the quality of background, we can add a suitable prompt (\emph{e.g.}, product photo, petals, water flowers \emph{etc.}), and the results can be better (three middle images). Finally, LoRA model is increased in generating process, the quality can be further improved (three lower images).

On the other hand, to distinguish the good and bad creatives in generating process, a reward model is proposed to predict the CTR, we only retain the good creatives with high CTRs to train LoRA model. At the same time, the good creatives with prompts generated by CLIP-Interrogator model\footnote{CLIP-Interrogator model is at \href{https://github.com/pharmapsychotic/clip-interrogator}{https://github.com/pharmapsychotic/clip-interrogator}.} are regarded as positive samples and the bad ones are regarded as negative samples to train prompt model. After multiple rounds of training processes, the LoRA model and prompt model can improve the quality of generated creatives and improve user experience.

\vspace{-4pt}
\subsection{Prompt Model}
The target of the prompt model is to select several appropriate words/tokens combined as one prompt, which can effectively describe the style of the image, and then input it into the stable diffusion model to generate more appealing images.

The input of the prompt model consists of four parts: user attributes (user age, user gender, \emph{etc.}), item attributes (item ID, category ID, \emph{etc.}), contextual attributes (user's device model, access time, \emph{etc.}), and tokens used for creative generation. The user attributes, item attributes, and contextual attributes are referred as \emph{Query}. The output of the prompt model is the estimated CTR score for each token in the whole word vocabulary\footnote{We have collected item images provided by sellers and designers and run CLIP-Interrogator to get the complete tokens, which are then post-processed, such as deleting stop-words and entity nouns, to form the whole word vocabulary.} under the given \emph{Query} conditions. The top-$p$ tokens with the highest CTRs are selected, which are assembled as a new prompt and it is then inputted into a stable diffusion model to generate creative images.

The network architecture of the prompt model is illustrated in sub-figure (b) of Fig. \ref{fig-pipeline}. On the \emph{Query} side, user attributes, item attributes, and contextual attributes are one-hot encodings and then converted into corresponding embeddings through lookup table. These three embeddings are concatenated into a query embedding named $emb_q$. Simultaneously, all tokens in the prompt generated from the creative image are multi-hot encodings and then transformed into multiple embeddings named $emb_p^*$ in the same way. Then we send $emb_p^*$ into transformers and a mean-pooling layer to obtain one final embedding $emb_p$ representing for the prompt. These two embeddings, $emb_q$ and $emb_p$ are concatenated and passed through DeepFM \cite{guo2017deepfm} to produce a final estimated score.

The prompts generated from the clicked creatives are regarded as positive samples, and other prompts from non-clicked creatives are treated as negative samples. This enables the model to select the appropriate prompts that are likely to improve CTR under \emph{Query} conditions.

We train the prompt model under the cross-entropy loss function using the clicked data, which we call this loss function as "hard loss" ($L_{hard}$ in Equation \ref{eq-pm-hard-loss}). Additionally, to reduce the exploration cost in creative generation, we utilize the output values from the reward model as the soft labels for auxiliary training in self-cycling in Fig \ref{fig-pipeline}. The final loss is a weighted sum of the hard loss and soft loss, where $\lambda_p$ is 0.1 in our experiments.
\vspace{-4pt}
\begin{small}
\begin{equation}
    p = PromptModel(Query, \; Prompt)
\end{equation}
\begin{equation}
\label{eq-pm-hard-loss}
    L_{hard} = CELoss(p, \; y)
\end{equation}
\begin{equation}
    L_{soft} = CELoss(p, \; y_{RM})
\end{equation}
\begin{equation}
    PromptLoss = (1-\lambda_p) \cdot L_{hard} + \lambda_p \cdot L_{soft} \label{eq-promptloss}
\end{equation}
\end{small}

Furthermore, individualized prompts can be created for different user groups based on user information, as different users may prefer creatives with different styles. The generated creatives can enhance user experience while reducing exploration costs. For example, without considering user information, the prompt model may select and combine some normal tokens, such as "minimalistic" and "translucent". But for a teenager, better tokens may be "electronic" and "translucent" while for an old user, "golden" and "minimalistic" may be more appropriate. The normal prompt combined with "minimalistic" and "translucent" might not be the favorite for either user group, hence failing to meet the actual needs.

% \begin{figure}[h]
%     \centering
%     \includegraphics[width=1\linewidth]{PM_structure.png}
%     \caption{Prompt Model Structure.}
%     \label{fig:enter-label}
% \end{figure}

\vspace{-5pt}
\subsection{Reward Model}
%The structure of reward model is show in sub-figure (c) of Fig. \ref{fig-pipeline}, which is used to predict the CTRs of the creatives from the same item generated by stable diffusion model. Assume that \emph{i}-th item $I_i$ has \emph{m} creatives, referred as ${c_1, c_2, ..., c_m}$. Each creative contains title, image and caption where title of creative is the same as the title of item. As the titles of all creatives from the same item are the same, the title feature seems redundant for predicting CTR score. To explain the title feature is still important for this task, we use multi-head self-attention sub-module to compute the relationship between the textual feature and visual feature of each creative. In details, the multi-head self-attention sub-module in reward model can learn the relationship (\emph{e.g.}, similarity) between textual feature and visual feature, such that the creative with high similarity score may be predicted with high CTR score. To make reward model to learn the creative content better, the caption information is also be used, which is acquired by CLIP-Interrogator model. CLIP-Interrogator model uses image as input and outputs the textual caption to describe the image. 
The structure of reward model is shown in sub-figure (c) of Fig. \ref{fig-pipeline}, which is used to predict the CTRs of the creatives from the same item generated by stable diffusion model. Each creative contains title, image and caption where title of creative is the same as the title of item. As the titles of all creatives are the same, the title feature seems redundant for predicting CTR scores. To explain why title feature is still important for this task, we use a multi-head self-attention sub-module to compute the relationship between the textual feature and visual feature of each creative. In detail, the multi-head self-attention sub-module can learn the relationship (\emph{e.g.}, similarity) between textual feature and visual feature, such that the creative with a high similarity score may be predicted with high CTR score. In addition, to make reward model to learn the creative content better, the caption information is also used, which is acquired by CLIP-Interrogator model. CLIP-Interrogator model uses the image as input and outputs the textual caption to describe the image. 

The title and caption information of \emph{j}-th creative for \emph{i}-th item are inputted into pre-trained BERT model \cite{devlin2018bert} to extract textual features, referred as $t_j$ and $c_j$, respectively (Equation \ref{eq-bert-tit-cap-img}). The pre-trained Swin model \cite{liu2021swin} is used to extract visual feature (referred as $i_j$) from image information (Equation \ref{eq-bert-tit-cap-img}). Both BERT model and Swin model are pre-trained in our commercial data with token mask loss and patch mask loss, respectively. In particular, BERT model is further fine-tuned in our commercial clicked data \cite{yang2023practice}. Two fully-connected layers are used to reduce the dimension of feature (Equation \ref{eq-fc-ttt}). To fusion the cross-model features and extract the relationship between textual and visual features, we use transformers \cite{vaswani2017attention} to get self-attention features (Equation \ref{eq-tran}). The fused title, image and caption features ($t_j^t$, $i_j^t$, $c_j^t$) are concatenated into final feature. Other two fully-connected layers are used to predict the CTR score ($p_j^i$) of \emph{j}-th creative for \emph{i}-th item (Equation \ref{eq-fc-ctr}).
% \begin{minipage}{0.22\textwidth}
\vspace{-4pt}
\begin{small}
\begin{equation}
    t_j = BERT(title_j), \; c_j = BERT(caption_j), \; i_j = Swin(image_j) \label{eq-bert-tit-cap-img}
\end{equation}
\begin{equation}
    t_j^f = FC_t(t_j), \quad c_j^f = FC_c(c_j), \quad i_j^f = FC_i(i_j) \label{eq-fc-ttt}
\end{equation}
\begin{equation}
    t_j^t, \; c_j^t, \; i_j^t = transformers(t_j^f, \; c_j^f, \; i_j^f) \label{eq-tran}
\end{equation}
\begin{equation}
    p_j^i = FC_{ctr}(concat(t_j^t, \; c_j^t, \; i_j^t)) \label{eq-fc-ctr}
\end{equation}
\end{small}
% \end{minipage}

%The target of reward model is to predict the CTR score of each creative. Then we need to calculate the real CTR value ($y_j^i$), as shown in Equation \ref{eq-real-ctr}, where $click_j^i$ and $impression_j^i$ means the number of clicks and impressions from \emph{j}-th creative for \emph{i}-th item. The problem of predicting the CTR of creative is much more difficult than predicting the CTR of item because creatives have not sufficient impression opportunities, directly estimate the real CTR value may lead to a high variance. To alleviate this problem, we use the empirical Bayes method \cite{wang2011click} to smoothen the CTR estimation. The smoothed CTR value ($\hat{y_j^i}$) can be calculated as shown in Equation \ref{eq-real-smo-ctr} where $\alpha$ and $\beta$ can be estimated through all historical data by the method \cite{wang2011click}. Compared the real CTR value, the smoothed CTR has lower variance and can benefit the training.
The target of reward model is to predict the CTR score of each creative. The real CTR value is calculated by the following equation. 
\vspace{-8pt}
\begin{small}
\begin{equation}
    \hat{y_j^i} = \frac{click_j^i}{impression_j^i} \label{eq-real-ctr}
\end{equation}
% \vspace{-5pt}
% \begin{equation}
%     \hat{y_j^i} = \frac{click_j^i + \alpha}{impression_j^i + \beta} \label{eq-real-smo-ctr} 
% \end{equation}
\end{small}

List-wise and point-wise loss functions are used to train reward model. The list-wise loss function is shown in Equation \ref{eq-listloss} where $\hat{y_j^i}$ and $p_j^i$ are the real CTR and predicted CTR, respectively. And $w_j^i$ is the sample impression weight of \emph{j}-th creative for \emph{i}-th item, shown in Equation \ref{eq-weight}. Through the list-wise loss function, the model focuses on ranking the creatives within the same item. Although the list-wise loss function can lead model to correctly rank the creatives, the absolute difference between the real and predicted scores may be very huge. Then we use point-wise loss function as an auxiliary function to enforce model to produce more accurate scores, as shown in Equation \ref{eq-pointloss}.
\vspace{-2pt}
\begin{small}
\begin{equation}
    L_{list} = \frac{\sum_{i} w^i \cdot L_{list}^i}{\sum_{i} w^i} \;, \quad L_{list}^i = -\sum_{j} \hat{y_j^i} \cdot log(p_j^i) \label{eq-listloss}
\end{equation}
\vspace{-10pt}
\begin{equation}
    w_j^i = \frac{impression_j^i}{\sum_{i} \sum_{j} impression_j^i} \;, \quad w^i = \sum_j w_j^i \label{eq-weight}
\end{equation}
\begin{equation}
    L_{point} = \frac{\sum_{i} w^i \cdot L_{point}^i}{\sum_{i} w^i} \;, \quad L_{point}^i = \sum_{j} (\hat{y_j^i} - p_j^i)^2 \label{eq-pointloss}
\end{equation}
\end{small}

As shown in Equation \ref{eq-loss}, the final loss is the sum of list-wise loss and point-wise loss, where $\lambda_r$ is a hyperparameter and is 0.1 in our experiments.
\begin{small}
\begin{equation}
    RewardLoss = (1-\lambda_r) \cdot L_{list} + \lambda_r \cdot L_{point} \label{eq-loss}
\end{equation}
\end{small}

% \vspace{-10pt}
\section{EXPERIMENTS}
In this section, we evaluate our proposed generation pipeline on online experiments. To validate the effectiveness of reward model, we use commercial data and public data compared with two other public methods. On the other hand, we make some ablation experiments for prompt model to explain the importance of considering user information. Finally, we perform some analyses and show some cases to demonstrate the effectiveness of the self-cycling training mode in our generation pipeline.

\vspace{-5pt}
\subsection{Online Results} We select five categories (including women shoes, women bags, travel, beauty and mobile) in our commercial scene for online experiments. For each item, the prompt model is used to select suitable tokens in the whole vocabulary as one appropriate prompt considering user features. As different users may have varying preferences for creatives, so we need to use prompt model to extract different prompts based on the characteristics of users. Then 10 creatives for each item are automatically generated by stable diffusion with LoRA model given the selected prompts. These creatives are then inputted into reward model to predict CTRs, and the top 5 creatives with highest predicted CTRs are retained. These retained creatives are regarded as training samples to further train LoRA model. To further optimize the quality of generated creatives in the next step, these retained creatives in current step are regarded as positive samples and others are regarded as negative samples, which are used to train prompt model. This can indirectly improve creative quality by facilitating better prompts in subsequent steps. With ten rounds of the above processes, top 5 creatives for each item are generated and uploaded to online platform. We use Epsilon-greedy \cite{watkins1989learning, franccois2018introduction} as online display strategy. %we use 30\% of online traffic to randomly select one creative for one impression and other 70\% to select the best creative with highest real CTR to display based on the historical data.

As shown in Table \ref{tab-online}, without the self-cycling training in our pipeline (referred as "w/o self-cycling" in the table), we only use the first versions of LoRA and prompt models to generate the creatives. The CTR and revenue improvements are only 4.2\% and 3.8\%, respectively. After using LoRA and prompt models with five rounds (referred as "Middle" in the table) and ten rounds (referred as "Ours" in the table) of training, the results become much better.

\begin{table}
\renewcommand\arraystretch{1.0}
\setlength{\abovecaptionskip}{0cm}
\setlength\tabcolsep{3pt}
  \caption{Online results.}
  \label{tab-online}
  \begin{threeparttable}
  \small
  \begin{tabular}{c|ccccccc}
    \hline
    Methods & \makecell{All} & \makecell{Women \\ Shoes} & \makecell{Women \\ Bags} & Travel & Beauty & Mobile \\
    \hline
    %Baseline & - & - & - & - & - \\
    %\makecell{same prompt\tnote{1}} & - & - & - & - & -  \\
    \makecell{w/o \\ self-cycling} & 4.2\tnote{1}/3.8\tnote{2} & 2.2/3.5 & 11.3/3.1 & 6.0/4.3 & 2.7/4.8 & 3.7/4.5  \\
    \makecell{Middle \tnote{3}} & 8.1/7.2 & 6.6/7.8 & 10.7/4.2 & 2.2/7.4 & 2.7/1.1 & 12.7/12.9  \\
    Ours & \textbf{10.4/9.7} & \textbf{7.0/9.0} & \textbf{12.9/4.3} & \textbf{9.9/12.6} & \textbf{11.8/7.9} & \textbf{14.2/14.5}  \\
  \bottomrule
\end{tabular}
\begin{tablenotes}
\footnotesize
\item[1]The first metric is CTR improvement compared with the baseline of using the original image with percentage.
\item[2]The second metric is revenue improvement with percentage.
\item[3]"Middle" means the middle versions of prompt model and LoRA model in self-cycling.
\end{tablenotes}
\end{threeparttable}
\vspace{-15pt}
\end{table}

\begin{table*}
\small
\setlength{\abovecaptionskip}{0cm}
\renewcommand\arraystretch{1.0}
  \caption{Ablation study on reward model.}
  \label{tab-rm}
  \begin{threeparttable}
  \begin{tabular}{cc|cccccc|cccc}
    \hline
    \multirow{2}*{Type} & \multirow{2}*{Details} & \multicolumn{6}{c}{Commercial data} &  \multicolumn{4}{c}{Public data} \\
    \cline{3-12}
    ~ & ~ & Top-1 $\uparrow$ & Top-2 $\uparrow$ & Top-3 $\uparrow$ & Top-4 $\uparrow$ & Top-5 $\uparrow$ & MSE $\downarrow$ & Top-1 $\uparrow$ & Top-2 $\uparrow$ & Top-3 $\uparrow$ & MSE $\downarrow$ \\ 
    \hline
    \multirow{2}*{Public methods} & VAM \cite{wang2021hybrid} & 18.01\% & 11.92\% & 6.59\% & 2.81\% & 1.72\% & 2.518 & 6.26\% & 3.80\% & 1.23\% & 3.544 \\
    ~ & Rank \cite{you2023image}\tnote{1} & 19.87\% & 10.68\% & 5.51\% & 3.07\% & 1.71\% & 2.615 & 7.32\% & 4.60\% & 1.70\% & 3.545 \\
    \hline
    \multirow{4}*{\makecell{Fuse input \\ information}} & Only image & 18.32\% & 13.62\% & 6.44\% & 3.22\% & 2.06\% & 2.314 & 9.07\% & 5.66\% & 2.23\% & 3.135 \\
    ~ & Image and title\tnote{2} & \underline{22.26\%} & 13.44\% & 7.21\% & 3.86\% & 2.52\% & 2.300 & 10.99\% & 6.91\% & 2.65\% & 3.094 \\
    ~ & Image and caption & 21.59\% & 12.99\% & 7.49\% & 3.69\% & 2.58\% & 2.404 & 9.91\% & 6.22\% & 2.31\% & 3.103 \\
    ~ & w/o transformers & 21.29\% & 11.32\% & 6.50\% & 2.90\% & 1.79\% & 2.145 & 11.20\% & 6.61\% & 2.56\% & \underline{3.057} \\
    % ~ & Ours & - & - \\
    \hline
    \multirow{3}*{\makecell{Pre-train}} & w/o pre-train & 21.05\% & 13.77\% & \underline{7.53\%} & 4.02\% & 2.51\% & 2.263 & 10.91\% & 6.64\% & 2.60\% & 3.250 \\
    ~ & Only pre-train BERT & 20.92\% & 13.36\% & \underline{7.53\%} & \underline{4.54\%} & \underline{3.20\%} & \underline{2.060} & 10.51\% & 6.44\% & 2.60\% & 3.433 \\
    ~ & Only pre-train Swin & 22.08\% & 13.75\% & 6.63\% & 3.98\% & 2.41\% & 2.198 & \underline{11.05\%} & 6.61\% & 2.64\% & 3.132 \\
    % ~ & Ours & - & - \\
    \hline
    \multirow{2}*{\makecell{Loss}} & Only list-wise & 21.91\% & 12.85\% & 5.69\% & 2.64\% & 2.51\% & 2.424 & 10.66\% & 6.53\% & 2.59\% & 3.231 \\
    ~ & Only point-wise & 21.35\% & \textbf{14.06\%} & 7.44\% & 3.86\% & 2.60\% & 2.111 & 10.81\% & \underline{6.64\%} & \underline{2.66\%} & 3.058 \\
    \hline
    - & Ours & \textbf{23.81\%} & \underline{13.84\%} & \textbf{7.56\%} & \textbf{4.79\%} & \textbf{3.29\%} & \textbf{2.052} & \textbf{11.29\%} & \textbf{7.00\%} & \textbf{2.76\%} & \textbf{3.047} \\
  \bottomrule
\end{tabular}
\begin{tablenotes}
\footnotesize
\item[1]As rank model uses entity textual knowledge and entity image knowledge while this information does not exist in the public data and commercial data, then we delete these input features to compare. 
\item[2]As there is no title information in public data, we use the averaged image features to simulate the title feature.
\end{tablenotes}
\end{threeparttable}
\vspace{-10pt}
\end{table*}

\vspace{-5pt}
\subsection{In-depth Analysis}
%We have performed two ablation studies on reward model and prompt model, respectively.
%\vspace{-2pt}
\subsubsection{Ablation Study on Reward Model.} As described in METHODS section, reward model plays a crucial role in the self-cycling training process of generating creative pipeline. In reward model, we use top-\emph{k} CTR uplift to evaluate the performance. The top-\emph{k} CTR uplift metric is calculated as follows: 
\vspace{-2pt}
\begin{small}
\begin{equation}
    CTR\text{-}uplift_k = \frac{CTR\text{-}score_k}{CTR\text{-}score_{base}}- 1 \label{eq-ctr-uplift}
\end{equation}
\begin{equation}
    CTR\text{-}score_k = \frac{\sum_i \sum_{j \in topk_iC} click_i^j}{\sum_i \sum_{j \in topk_iC} impression_i^j} \label{eq-ctr-score}
\end{equation}
\begin{equation}
    CTR\text{-}score_{base} = \frac{\sum_i \sum_{j \in C} click_i^j}{\sum_i \sum_{j \in C} impression_i^j} \label{eq-ctr-score-base}
\end{equation}
\end{small}
, where $CTR\text{-}score_{k}$ is the cumulative CTR scores of top-\emph{k} creatives with the highest scores predicted by reward model for each item. $click_i^j$ and $impression_i^j$ are the numbers of clicks and impressions of \emph{j}-th creative for \emph{i}-th item. $topk_iC$ is the set of top-\emph{k} creatives for the \emph{i}-th item. If $CTR\text{-}uplift_k$ metric is higher, the performance of the reward model is better. The high top-$k$ CTR uplift metrics only ensure that the order of outputs is good, so we use another MSE metric to measure the difference between the predicted CTRs and real ones as follows:
\vspace{-5pt}
\begin{small}
\begin{equation}
    MSE = \frac{1}{n} \sum_{i=1}^{n}  \frac{1}{m_i} \sum_{j=1}^{m_i} (\hat{y_j^i} - p_j^i)^2 \label{eq-mse}
\end{equation}
\end{small}
, where $n$ is the number of items and $m_i$ is the number of creatives in $i$-th item.

The commercial data and public data\footnote{The public data is at \href{https://tianchi.aliyun.com/dataset/93585}{https://tianchi.aliyun.com/dataset/93585}.} \cite{wang2021hybrid} are tested to demonstrate the effectiveness of our proposed reward model structure. The commercial data has 1.0M creatives from 286k items. Each creative has item title, image, caption (generated by CLIP-Interrogator), real click and real impression. Creatives from the 5\% of items are randomly sampled as the testing data while the others are training data. The public data contains 1.7M creatives from 500k items. As there is no title information in public data, we average the image features of creatives from the same item as the title feature.

The BERT and Swin in reward model are pre-trained on text information (title and caption) and image information with token mask loss and patch mask loss, respectively. Then in training reward model process, the parameters of BERT and Swin are fixed to speed up. We use adam optimizer with mini-batch of 2048 and learning rate of 5e-4. The number of epochs is set 30.

We tested two public methods, including VAM \cite{wang2021hybrid} and Rank \cite{you2023image}. On the commercial data, the top-1 to top-5 CTR uplift scores of our proposed method are much higher than VAM and Rank while the MSE metric of the proposed method is smaller than other methods. We have similar results on the public data. In the reward model, we use title, image and caption information to predict the CTR scores. To demonstrate that the title and caption information is also helpful for this task, this information is deleted in the original network. As shown in Table \ref{tab-rm}, only considering the image feature to predict the CTR, the results are not good. Simply concatenating three multi-modal features without transformers can improve the CTR ranking effect, but the improvement is limited. Fusing transformers can further improve the results, which verifies the effectiveness of combining different types of input information and using transformers to extract the multi-model feature.

In our proposed method, we have pre-trained BERT and Swin models with mask loss on corresponding data. As shown in this table, the results without pre-training are slightly inferior to our results, which verifies the importance of pre-training. Lastly, we have tested the effects of the loss functions. The results show that both list-wise and point-wise loss functions are useful for improving the results.

\begin{table}
\setlength{\abovecaptionskip}{0cm}
\renewcommand\arraystretch{1.0}
  \caption{Ablation study on prompt model.}
  \label{tab-pm-exp}
  \begin{threeparttable}
  \small
  \begin{tabular}{cc|ccc}
    \hline
    Type & Details & Historical CTR\tnote{1} & CTR & Revenue \\
    \hline
    \multirow{2}*{User ablation} & w/o user & 9.4\%& 7.2\%& 6.9\%\\
    ~ & Individuation & 12.5\%& 10.4\%& 9.7\%\\
    \hline
    % \multirow{2}*{reward model} & w/o & - & - & - \\
    % ~ & with & - & - & - \\
    % \hline
    % \multirow{2}*{self-cycling} & start & - & - & - \\
    % ~ & final & - & - & - \\
    % \hline
    \multirow{3}*{\makecell{Model \\ structure}} & w/o transformer & 10.9\% & -\tnote{2} & - \\
    ~ & w/o DeepFM & 11.7\% & - & - \\
    ~ & Ours & 12.5\% & - & - \\
  \bottomrule
\end{tabular}
\begin{tablenotes}
\footnotesize
\item[1]This is an offline metric, it is the historical CTR improvement given the clicked data. In detail, we can calculate the cumulative CTR improvements of top-$k$ tokens.
\item[2]Model structure experiments are not performed online, and so only the offline metric of "Historical CTR" is shown.
\end{tablenotes}
\end{threeparttable}
\vspace{-10pt}
\end{table}

% \vspace{-10pt}
\begin{figure}[h]
\setlength{\abovecaptionskip}{0.0cm}
  \centering
  \includegraphics[width=1.0\linewidth]{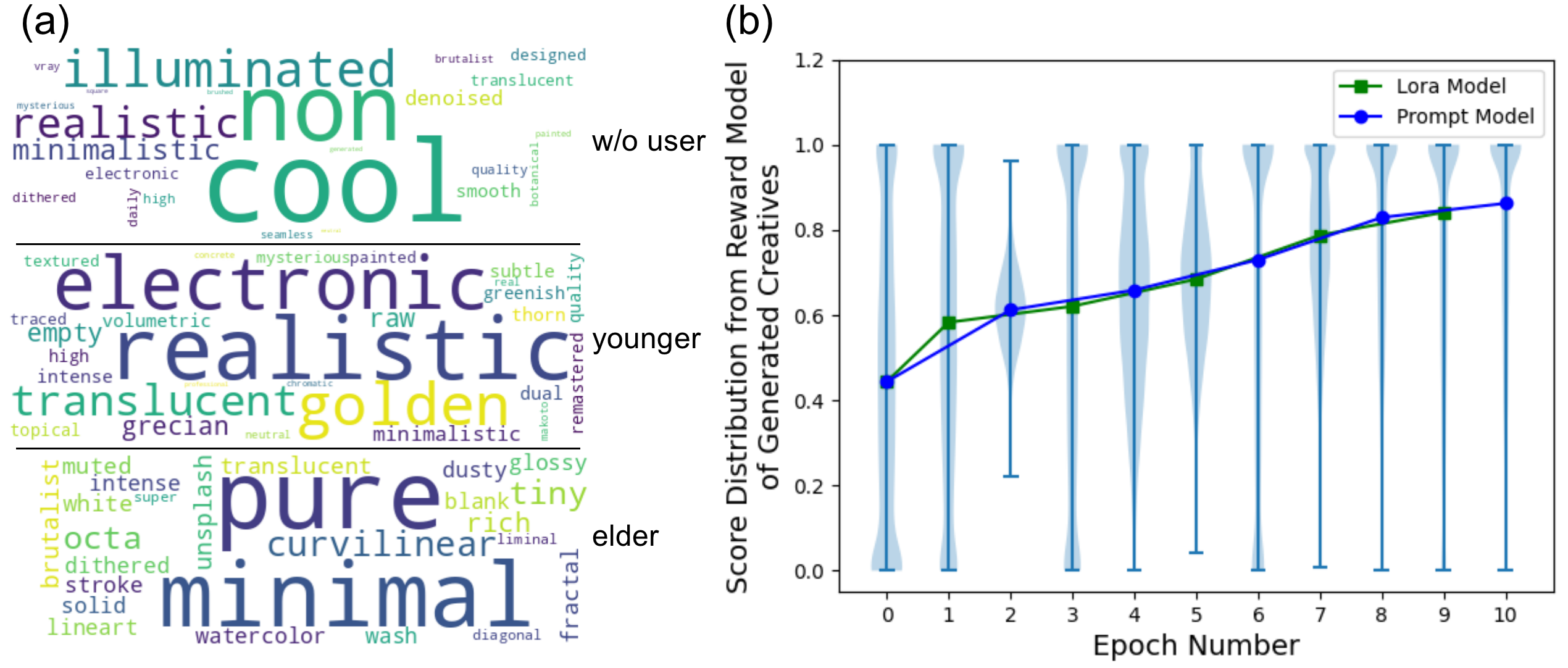}
  \caption{(a) Word cloud analysis of generated tokens by prompt models without and with considering user information. (b) The creatives generated by different versions of LoRA and prompt models are scored by reward model to show the effectiveness of self-cycling training process.}
  \label{fig-exp-self}
  \Description{(a) Word cloud analysis of generated tokens by prompt models without and with considering user information. (b) The creatives generated by different versions of LoRA and prompt models are scored by reward model to show the effectiveness of self-cycling training process.}
  \vspace{-15pt}
\end{figure}

% \begin{table}
% \renewcommand\arraystretch{1.0}
%   \caption{Ablation study on reward model for ranking strategy.}
%   \label{tab-rm-rank}
%   \begin{threeparttable}
%   \begin{tabular}{ccc}
%     \hline
%     \makecell{Ranking Strategy \\ on Public Data} & \makecell{whether to \\ add the reward model} & sCTR (\%) \\
%     \hline
%     \multirow{2}*{E-Greedy} & no & - \\
%     ~ & yes & - \\
%     \hline
%     \multirow{2}*{Thompson Sampling} & no & - \\
%     ~ & yes & - \\
%     \hline
%     \multirow{2}*{UCB} & no & - \\
%     ~ & yes & - \\
%   \bottomrule
% \end{tabular}
% \end{threeparttable}
% \end{table}

%Similar with the method in [ref], we also tested the performance of features extracted by the reward model on online creative ranking strategy. As shown in Table \ref{tab-rm-rank}, three different strategies, \emph{e.g.}, E-Greedy [ref], Thompson Sampling [ref], UCB [ref], are tested on the public data without or with features from the reward model. The metric sCTR in this table is the simulated CTR, which is used to simulate the real online CTR. With considering the features, all creative ranking strategies have the steady improvements.

\vspace{-8pt}
\begin{figure*}
\setlength{\abovecaptionskip}{-0.2cm}
\setlength{\belowcaptionskip}{-0.4cm}
  \centering
  \includegraphics[width=0.75\linewidth]{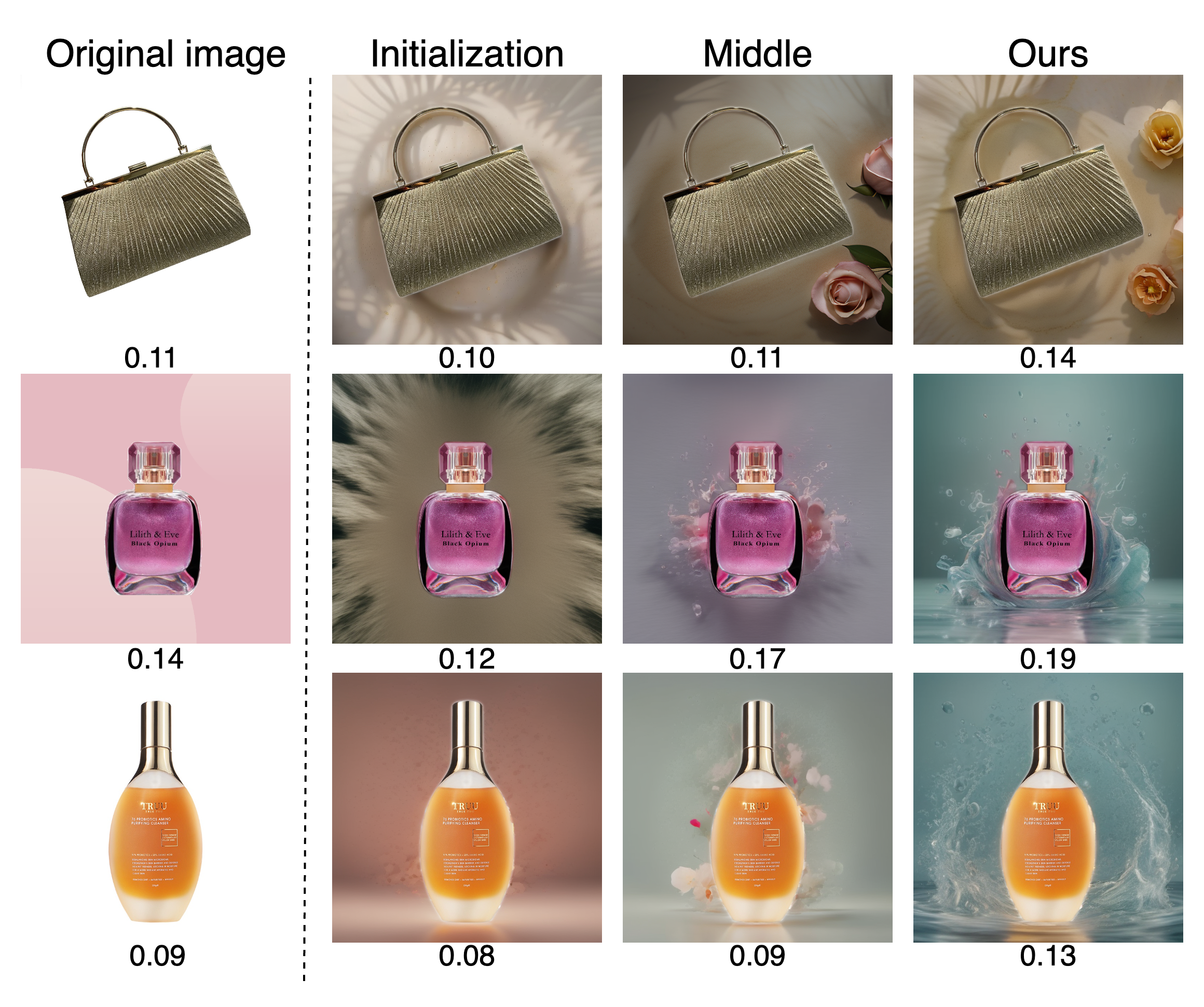}
  \caption{Cases with the original image and generated creatives in different steps of self-cycling. Bottom score is the real CTR.}
  \label{fig-show-cases}
  \Description{Different cases with the original image and generated creatives in different steps of self-cycling training process.}
  \vspace{-2pt}
\end{figure*}

% \vspace{-10pt}
\subsubsection{Ablation Study on Prompt Model.} This section will show the ablation study of the prompt model. We use online clicked data to train the prompt model, with 20M samples as training data and 2M samples as testing data. Each sample has an average of 8.5 tokens. We employ Adam as the optimizer, set the learning rate to 1e-3, and train the model in 2 epochs to avoid overfitting with a batch size of 2048. The parameter of prompt model trained in this clicked data is initialized as the first version of prompt model in self-cycling training process.

%We perform two experiments to severally explain the advantage of importing the user groups information and the improvements of prompt model in self-cycling training process. In the importing the user groups information experiment, the user groups information is deleted in prompt model in both training and testing processes, the CTR and revenue improvements are both limited. To further improve the results, we import the user groups information. In details, for each item, all user groups are enumerated and select one best prompt for each user group. Assuming that there are 5 creatives for each item without user information, after introducing user features, there are $5 \times u$ creatives where $u$ refers to the number of user groups. In online display strategy, we select the corresponding 5 creatives with the same user group as the current user for personalized creative recommendation. As shown in this table, the metrics in "individuation" have great improvements. 
We perform two experiments to severally explain the advantage of importing the user group information and the improvements of prompt model in self-cycling training process. In the fusing user group information experiment, the user group information is deleted in prompt model in both training (in self-cycling training mode) and testing processes, the CTR and revenue improvements are both limited (Table \ref{tab-pm-exp}). To further improve the results, we import the user groups information. In detail, for each item, all user groups are enumerated, and one best prompt is selected for each user group by prompt model. Assuming that there are 5 creatives for each item without user information, after introducing user features, there are $5 \times u$ creatives where $u$ refers to the number of user groups. In online display strategy, we select the corresponding 5 creatives with the same user group as the current user for personalized creative recommendation. As shown in this table, the metrics in "individuation" have great improvements. In addition, we collected the tokens generated by prompt model to perform the word cloud analysis (Fig. \ref{fig-exp-self}(a)). After considering the user group information (\emph{e.g.}, younger and elder), the generated tokens are more "individualized" where younger users prefer "electronic" and "realistic" while elder users prefer "pure" and "minimal". This phenomenon illustrates that prompt model can learn the fact that different user groups have various preferences for diverse tokens.

In another experiment of self-cycling training, we compare the results of the first version and last version of prompt model (Table \ref{tab-online}). After self-cycling training, the final version of prompt model has higher metrics which further verifies the necessity of self-cycling. Lastly, we find that the transformers and DeepFM sub-modules in prompt model are also helpful with good historical CTR improvements.

\vspace{-6pt}
\subsubsection{Self-cycling Analysis}
In previous section, we have already demonstrated the effectiveness of reward model on commercial and public data. Then we can use reward model to offline analyze the performances of LoRA and prompt models in self-cycling training process. As shown in Fig. \ref{fig-exp-self}(b), the reward scores of generated creatives increase during self-cycling training, demonstrating the effectiveness of our proposed training process.

% \vspace{-10pt}
\subsection{Case Study} Lastly, we will show some amazing cases to validate the effectiveness of our proposed creative generation pipeline (Fig. \ref{fig-show-cases}). The aesthetics and harmony of creatives generated by our proposed method are much better than the original image provided by seller.

% \vspace{-10pt}
\section{DISCUSSION}
A new automated \textbf{C}reative \textbf{G}eneration pipeline for \textbf{C}lick-\textbf{T}hrough \textbf{R}ate (CG4CTR) is proposed to fuse the CTR target into the creative generation task. In CG4CTR, we first use inpainting mode in diffusion method to generate the creative image in advertising scene. The generation process is carried out by self-cycling, where the prompt model and LoRA model are updated alternately and iteratively. As the traditional creative generation methods do not use the CTR as the target, some bad creatives with poor CTR performance may be generated, leading to sub-optimal online results. To solve this problem, we fuse the CTR target in CG4CTR and consider user information in prompt model to select the appropriate prompt and indirectly generate individualized creatives. In self-cycling, we need the reward model to judge which creatives are good and which ones are bad, and then it can help to further improve the quality of generated creatives. Both offline and online experiments show that CG4CTR can generate better creatives with higher CTR compared with the original images provided by sellers.

Our proposed CG4CTR pipeline can be inserted at different stages of the item recommendation. Specifically, if CG4CTR is inserted after the item recommendation (named A-stage), for one user comes, candidate items are ranked by the ranking model, then the top items are used for CG4CTR to generate creatives for each item. In A-stage, the ranking model can only consider the original image rather than creative image, leading to the reduced performance. If CG4CTR is inserted before the item recommendation (named B-stage), the creatives for all candidate items are generated by CG4CTR, then ranking model can simultaneously consider both item information and creative information to sort them, which may have better performance but also need higher online time consuming. In future work, we will explore the difference in performance between these two stages. Furthermore, we can employ various styles of LoRA models (rather than one LoRA in CG4CTR) to generate creatives with enhanced styles and superior quality, where these styles can be provided by designers as initial samples to train LoRA model, and then their parameters are updated in self-cycling training mode. Lastly, more ways to generate creatives rather than modifying the background need to be investigated in the same framework.

\bibliographystyle{ACM-Reference-Format}
\balance
\bibliography{paper_list}
\end{document}